\documentstyle[12pt]{article}
\date{}
\begin{document}
\title {\Large\bf Lauglin-type wavefunction of two-dimensional electrons in 
          the tilted magnetic field}
\author{\normalsize Shi-Jie Yang$^1$, Yue Yu$^2$, Jin-Bin Li$^2$}
\maketitle

{\small\it
 1. Center for Advanced Study, Tsinghua University, Beijing 100084, China \\
         2. Institute of Theoretical Physics, 
Chinese Academy of Sciences, P. O. Box 2735, Beijing 100080, China}

\vspace{1mm} 
\begin{abstract} 
We study the fractional quantum Hall states in 
the tilted magnetic field. A many-particle wavefunction of the ground state , 
which is similar 
to that of Laughlin's, is constructed in the Landau gauge. We show that in the 
limit of thermodynamics, the concept 
of composite fermion is still valid in presence of the in-plane field.

\end{abstract}
\vspace{4mm}
\hspace{2cm}{PACS number(s): 73.20.Dx, 73.40.Hm, 73.40.Kp, 73.50.Jt}
\vskip 2pc

Recently, two-dimensional electron system (2DES) in tilted magnetic field has 
attracted great interests in both experimentalists and theorists.
The magneto-transport experiments on high mobility samples in GaAs/AlGaAs 
heterostructures revealed new classes of correlated many-electron states 
\cite{Eisenstein1}. The most prominent findings are the discoveries of the 
giant anisotropy in the resistivity near half filling of the topmost LL\cite{Lilly,Du}. 
It is revealed that anisotropy occurs when the 2DES is applied by an in-plane 
magnetic field. The easy direction of transport is perpendicular to the in-plane field.  
It is generally accepted that the highly anisotropic transport is related to the 
formation of the unidirectional charge-density-wave (UCDW) state, {\it i.e.}, 
the stripe phase \cite{Jungwirth,Phillips}. The possibility of existence 
of the UCDW was originally predicted by Koulakov, Fogler, and Shklovskii\cite{Koulakov} 
based on a little earlier work on the Hartree-Fock treatment of 
the high Landau levels. Specifically, the extotic
$\nu=5/2$ fractional quantum Hall state, which shows no anisotropy in perpendicular 
magnetic field, becomes highly anisotropic when the external field is tilted an angle.
Contrary to other odd-denominator filling states, which occur as Jain' series, 
the Hall plateau of the incompressible 
$\nu=5/2$ state is explained as the appearance of ground state of spin-singlet 
pairing of composite fermions (CFs)\cite{HaRe}. However, the spin-polarized $p$-wave 
BCS paring of CFs, or the Moore-Read(MR) Pfaffian wave function\cite{MR}, may be 
another possibility \cite{GWW}, which was recently suggested to be favorable 
\cite{morf,RH}. Studies by Eisenstein et al \cite{Eis} in the tilted field experiments 
have shown that the plateau disappears if the tilted angle $\theta$ exceeds a 
critical value. The explanation of the experiments from the point of view of 
the singlet-pairing can be understood as a gain in Zeeman energy \cite{Eis1}.
However, the Lande $g$-factor is about $30\%$ larger than expected.
On the other hand, in the picture of $p$-wave paring of CFs, how the tilted field 
violates 
the spin-polarized paired Hall state is still puzzling. Yu {\it et al}\cite{Yu}
proposed a mechanism to solve the above puzzle. They considered that there exists 
a competition between the instabilities of the CF Fermi surface to the formation 
of the UCDW and the paired Hall state. When the tilted angle is small, the pairing 
state dominates. But as the tilted angle increases, the UCDW takes energetically 
over the paired Hall state as the ground state, which transforms the incompressible 
state to the compressible 
state. Recent experiment by Pan {\it et al}\cite{Pan} supports their suggestion.

Here comes the question. One may ask if the concept of the composite fermion is still valid
when the magnetic field is tilted an angle. In the present work, we prove that in the 
limit of thermodynamics, one can construct a Laughlin-type wavefunction based on a similar
reasoning of the original work by Laughlin. The concept of composite fermion in tilted
field can be deduced from the analysis of the Laughlin-type wavefunction in an analogous
way taken by J. K. Jain\cite{Jain}.

Consider an electron moving in a torus geometry under the influence of a strong 
magnetic field which is tilted an angle $\theta$ to 
the $x-y$ plane, with $B_x=B \tan \theta$ and $B_z=B$. The electron is confined in a 
harmonic potential $V(z)={1\over 2}m_b \Omega^2 z^2$ in the $z$-direction, where $m_b$
is the band mass of the electron. If the characteristic frenquence $\Omega\gg\omega_c$,
where $\omega_c={\hbar\over m_b l_0^2}$ is the electron cyclotron frenquence 
in the perpendicular magnetic field $B$, then the electrons move in a quasi-two-
dimensional plane. We work in the 
"Landau gauge" by choosing the vector potential ${\bf A}=\{0, x B_z-z B_x, 0\}$. The 
single-electron Hamiltonian is then,
\begin{equation}
\hat{H}={1\over 2m_b}\left [(-i\hbar\partial_x)^2+
         (-i\hbar\partial_y-{e\over c}(x B_z-z B_x))^2 
         +(-i\hbar \partial_z)^2\right]
       +{1\over 2}m_b \Omega^2 z^2.
\end{equation}

Take the length unit $l_0^2=\hbar c/e B=1$ and seperate out the plane-wave in the 
$y$-direction,
\begin{equation}
\Psi(x,y,z)={1\over \sqrt{L_y}} e^{i k y} \phi (x,z),
\end{equation}
where $k=2\pi j/L_y=k_0\cdot j$, ($j=1,2,\cdots$). Then the Hamiltonian can be 
equivalently rewritten as
\begin{equation}
\hat{H}={1\over 2}\hbar \omega_c \left [-\partial_\xi^2-\partial_z^2
    +\xi^2-2\xi z\tan \theta+(\tan^2\theta+{\Omega^2\over \omega_c^2})z^2\right],
\end{equation}
where $\xi=x-x_0$ with $x_0=kl_0$.

To decouple the two coupled harmonic oscillators, we make a coordinate-rotation 
in the $\xi-z$ plane:
\begin{equation}
 \left \{\begin{array}{ll}
\xi=&u\cos \alpha-v \sin\alpha \\
z=&u\sin\alpha+v\cos\alpha
\end{array}
\right.
\end{equation}
Take $\tan\alpha=\left[{\omega_c^2\over (\omega_+^2-\omega_c^2)}\right]\tan\theta$,
then the Hamiltonian becomes
\begin{equation}
\hat{H}={1\over 2}\hbar \omega_{-}(-{\partial^2\over \partial \zeta^2}+\zeta^2)
+{1\over 2}\hbar \omega_{+}(-{\partial^2\over \partial \eta^2}+\eta^2).
\end{equation}
where $\zeta={l_0\over l_{-}}u$ and $\eta={l_0\over l_+}v$ 
with $l_{\pm}^2=\hbar/m_b\omega_{\pm}$, and
\begin{equation}
\omega_{\pm}^2={1\over 2}(\Omega^2+{\omega_c^2\over \cos^2\theta})\pm 
{1\over 2}\sqrt{(\Omega^2-{\omega_c^2\over \cos^2\theta})^2
+4\Omega^2 \omega_c^2 \tan^2\theta}.
\end{equation}
Hence, for $\omega_{\pm}$, the corresponding eigen wavefunctions are
\begin{eqnarray}
\phi_{n}^{\omega_{+}}&=&N_{n}^{+}H_{n}\left({-(x-x_0)\sin\alpha
                        +z\cos\alpha\over l_{+}}\right)
                    \cdot e^{-[-(x-x_0)\sin\alpha+z\cos\alpha]^2/2l_{+}^2}, \\
\phi_{n}^{\omega_{-}}&=&N_{n}^{-}H_{n}\left({(x-x_0)\cos\alpha
                        +z\sin\alpha\over l_{-}}\right)
                     \cdot e^{-[(x-x_0)\cos\alpha+z\sin\alpha]^2/2l_{-}^2},
\end{eqnarray}
where $H_{n}(x)$ is the Hermitian polynomials and 
$N_n^{\pm}={1\over \sqrt{2^n n!\sqrt{\pi}l_{\pm}}}$ is the normalization coefficients.

As we are concerning the lowest Landau level, the degenerate single-particle 
wave functions are
\begin{eqnarray}
\Psi_j&=&{e^{iky}\over \sqrt{L_y}}{1\over \sqrt{\pi l_{-}l_{+}}} e^{-[(x-x_0)\cos\alpha
   +z\sin\alpha]^2/2l_{-}^2-[-(x-x_0)\sin\alpha+z\cos\alpha]^2/2l_{+}^2} \nonumber \\
      &\propto&e^{-{1\over2} j^2[{l_0^4\over l_{-}^2} k_0^2\cos^2\alpha
   +{l_0^4\over l_{+}^2} k_0^2\sin^2\alpha]}\cdot e^{j[ik_0 y
   +{l_0^2\over l_{-}^2}k_0\cos\alpha (x\cos\alpha+z\sin\alpha)
   -{l_0^2\over l_{+}^2} k_0 \sin\alpha (-x\sin\alpha+z\cos\alpha)]} \nonumber \\
      & &\cdot e^{-{1\over 2 l_{-}^2}(x\cos\alpha+z\sin\alpha)^2
   -{1\over 2 l_{+}^2}(-x\sin\alpha+z\cos\alpha)^2} \nonumber \\
      &=&e^{-{1\over2} j^2[{l_0^4\over l_{-}^2} k_0^2\cos^2\alpha
   +{l_0^4\over l_{+}^2} k_0^2\sin^2\alpha]}\cdot e^{ju-v^2}.
\end{eqnarray}
with
\begin{eqnarray}
u&=&ik_0 y+{l_0^2\over l_{-}^2}k_0\cos\alpha (x\cos\alpha+z\sin\alpha)
-{l_0^2\over l_{+}^2} k_0 \sin\alpha (-x\sin\alpha+z\cos\alpha) \\
v^2&=&{1\over 2 l_{-}^2}(x\cos\alpha+z\sin\alpha)^2
-{1\over 2 l_{+}^2}(-x\sin\alpha+z\cos\alpha)^2,
\end{eqnarray}

Since the wavefunction is localized around $x_0$ in the $x$-direction, the edge effect 
can be omitted in the limit of thermodynamics. The many-particle wavefunction 
for the filled lowest LL is expressed in the Slater determinant form,
\begin{eqnarray}
\Phi ({\bf r}_1,{\bf r}_2,\cdots,{\bf r}_N)&=&\left | \begin{array}{ccc}
  \Psi_1({\bf r}_1) & \Psi_1({\bf r}_2) & \cdots \\
  \Psi_2({\bf r}_1) & \Psi_2({\bf r}_2) &        \\
  \vdots            &                   &        
\end{array} \right |  \nonumber \\
&\propto&\left | \begin{array}{ccc}
  e^{u_1} & e^{u_2} & \cdots \\
  e^{2u_1}& e^{2u_2}&        \\
  \vdots  &         &           
\end{array} \right |\cdot e^{-\sum_k v_k^2} \nonumber \\
&=&\prod_{j<k}(e^{u_j}-e^{u_k})e^{\sum_k (ik_0 y_k-w_k^2)},
\end{eqnarray}
which simply reduces to a Vandermonde determinant. Here we denote
\begin{equation}
w^2\equiv [(x-k_0 l_0^2)\cos\alpha+z\sin\alpha]^2/2l_{-}^2
+[-(x-k_0 l_0^2)\sin\alpha+z\cos\alpha]^2/2l_{+}^2 .
\end{equation}

Till now our wavefunction is equivalent to that of in the symmetric gauge except 
an unimportant phase factor.
Let us now extend it to the fractional filling states $\nu={1\over 2p+1}$. In analogy 
to Laughlin's analysis in the symmetric gauge in absence of in-plane field, 
the wavefunction must satisfy the following conditions:\\
a. It must be anti-symmetric for exchanging any two electrons;\\
b. The state $|m\rangle$ is the eigenstate of the momentum in $y$-direction.

One finds that the unique form of the wavefunction for FQHE state $\nu=1/m$ 
($m=2p+1$ is an odd number) is
\begin{equation}
\Phi_{1/m}=\prod_{j<k}(e^{u_j}-e^{u_k})^m e^{imk_0 \sum_k y_k}e^{-\sum_{k}w_k^2}
\label{wave}
\end{equation}
The total momentum in $y$-direction is $K={1\over 2}N(N+1)\cdot mk_0$ with $N$ 
the number of electrons.
\begin{equation}
\hat{P}_y \Phi_{1/m}=K\Phi_{1/m}
\end{equation}

According to J. K. Jain\cite{Jain}, we rewrite the Eq.(\ref{wave}) as
\begin{eqnarray}
\Phi_{1/m}&=&\prod_{j<k}(e^{u_j}-e^{u_k})e^{ik_0\sum_k y_k}\cdot e^{-\sum_k w_k^2}
    \cdot \prod_{j<k}(e^{u_j}-e^{u_k})^{2p} e^{2pik_0 \sum_k y_k} \nonumber \\
      &=& \Phi_1\cdot \chi_0^{2p}.
\end{eqnarray}
Here $\Phi_1$ is the wavefunction for one filled Landau level. 
This Jastrow-form wavefunction can be considered as each electron carry $2p$ 
flux quanta, therefore,
we recover Jain's concept of "composite fermions" in the tilted field. 
In this picture, the electrons "nucleate" even number of flux to screen enough of the 
external magnetic field, so that the composite fermions exactly fill an integer number
of Landau levels associated with the surplus part of unscreen field 
($B^*=B-2p\phi_0\bar{\rho}$, with $\phi_0=hc/e$ the unit magnetic quanta 
and $\bar{\rho}$ the average particle density).
The wavefunctions of general fractional filling factors ($\nu=n/2pn\pm 1$) are
explicitly expressed as 
\begin{equation}
\Phi_{n/2pn\pm 1}={\cal P}_{LLL} \Phi_{n}^{*}\cdot\chi_0^{2p},
\end{equation}
where $\Phi_n^*$ is the wavefunction for composite fermions filling $n$ LLs. 
The operator ${\cal P}_{LLL}$ projects the wavefunction onto the lowest Landau level.

In summary, we have written a many-particle wavefunction for fractional quantum 
Hall states with in-plane magnetic field. We worked in the Landau gauge 
because in presence 
of an in-plane field, where the relative angular momentum between two electrons is not 
a good quantum number. Haldane\cite{Haldane} had written 
a many-particle wavefunction in the torus geometry with the magnetic field
perpendicular to the 2DES plane. 
It is difficult to reproduce Laughlin's wavefunction in the symmetric gauge in 	
presence of an in-plane field. Our result is obtained by the same reasoning
employed by Laughlin. It shows that fractional quantum Hall
states in the lowest Landau level survive even when the 2DES is applied by an in-plane
field. Hall plateaus at these filling factors can be observed in experiments. We 
conclude that the concept of composite fermion in tilted field is still valid by 
a way analogous to that of J. K. Jain. It should be 
noted that the explicit form of composite fermion is different from that of 
Jain's\cite{Jain}. Our result provides a supplementary proof 
to the explanation for the $\nu=5/2$ state when the external magnetic field is tilted 
an angle, where the competition between paired state of composite fermions and
unidirectional charge-density-wave state leads to the destroy of the pairing gap and the 
anisotropic transport subsequently takes place\cite{Yu}.

This work is supported in part by the NSF of China.


\vspace{-0.1in}

\begin{thebibliography}{99}

\bibitem{Eisenstein1}  For a review, see J. P. Eisenstein,  M. P. Lilly, K.
B. Cooper,  L. N. Pfeiffer, and K. W. West, cond-mat/9909238.

\bibitem{Lilly} M. P. Lilly, K. B. Cooper, J. P. Eisenstein,
L. N. Pfeiffer and K. W. West, Phys. Rev. Lett. {\bf 82}, 394
(1999).

\bibitem{Du}  R. R. Du, D. C. Tsui, H. L. Stormer, L. N. Pfeiffer, K. W.
Baldwin, and K. W. West, Solid Stat. Commun. {\bf 109}, 389(1999).

\bibitem{Jungwirth}  T. Jungwirth, A. H. MacDonald, L. Smrcka and S. M. Girvin,
Phys. Rev. B {\bf 60}, 15574 (1999).

\bibitem{Phillips}  T. Stanesca, I. Martin and P. Phillips, 
Phys. Rev. Lett. {\bf 84}, 1288 (2000).

\bibitem{Koulakov}  A. A. Koulakov, M. M. Fogler, and B. I. Shklovskii, 
Phys. Rev. Lett. {\bf 76}, 499(1996); M. M. Fogler,  A. A. Koulakov, and B.
I. Shklovskii, Phys. Rev. B {\bf 54},  1853(1996).

\bibitem{HaRe}  F. D. M. Haldane and E. H. Rezayi, Phys. Rev. Lett. 
{\bf 60}, 956 (1988).

\bibitem{MR}  G. Moore and N. Read, Nucl. Phys. B{\bf 360}, 362, (1991).

\bibitem{GWW}  M. Greiter, X. G. Wen and F. Wilczek, Phys. Rev. Lett. 
{\bf 66}, 3205 (1991); Nucl. Phys. B {\bf 374}, 567 (1992). K. Park, 
V. Melik-Alaverdian, N. E. Bonesteel and J. K. Jain, Phys. Rev. B {\bf 58},
10167 (1998). N. E. Bonesteel, Phys. Rev. Lett. {\bf 82}, 984 (1999).

\bibitem{morf}  R. H. Morf, Phys. Rev. Lett. {\bf 80}, 1505 (1998).

\bibitem{RH}  E. H. Rezayi and F. D. M. Haldane, cond-mat/9906137.

\bibitem{Eis}  J. P. Eisenstein, R. L. Willett, H. L. Stormer, D. C. Tsui,
A. C. Gossard and J. H. English, Phys. Rev. Lett. {\bf 61}, 997 (1988).

\bibitem{Eis1}  J. P. Eistenstein, R. L. Willett, H. L. Stormer, L. N.
Pfeiffer and K. W. West, Surf. Sci. {\bf 229}, 31 (1990).

\bibitem{Yu}  Yue Yu, Shi-Jie Yang, and Zhao-Bin Su, Phys. Rev. B {\bf 62},
15371(2000).

\bibitem{Pan} W. Pan, H.L. Stormer, D.C. Tsui, L.N. Pfeiffer, K.W. Baldwin, 
K.W. West, cond-mat/0103144.

\bibitem{Jain}  J. K. Jain, Phys. Rev. Lett. {\bf 63}, 1223(1989); Phys.
Rev. B {\bf 41}, 7653(1990).

\bibitem{Haldane} F.D.M.Haldane, Phys.Rev.Lett.{\bf 55}, 2095 (1985).

\end{thebibliography}
\end{document}